\documentclass[preprint]{aastex}
\shortauthors{Li et al.}
\shorttitle{Dynamical-Chemical Model of Molecular Cloud Cores}

\begin{document}

\title{A Coupled Dynamical and Chemical Model of Starless Cores of 
\break Magnetized Molecular Clouds: I. Formulation and Initial Results}

\author{Zhi-Yun Li}
\affil{Department of Astronomy, University of Virginia, P.O. Box 3818,
Charlottesville, VA 22903}
\and
\author{V. I. Shematovich, D. S. Wiebe and B. M. Shustov}
\affil{Institute of Astronomy of the RAS, 48, Pyatnitskaya str.,
109017 Moscow, Russia}

\begin{abstract}
We develop a detailed chemical model for the starless cores of strongly
magnetized molecular clouds, with the ambipolar diffusion-driven dynamic
evolution of the clouds coupled to the chemistry through ion
abundances. We concentrate on two representative model clouds in this
initial study, one with magnetic fields and the other without. The
model predictions on the peak values and spatial distributions of
the column densities of CO, CCS, N$_2$H$^+$ and HCO$^+$ are compared
with those observationally inferred for the well-studied starless core
L1544, which is thought to be on the verge of star formation. We find
that the magnetic model, in which the cloud is magnetically supported
for several million years before collapsing dynamically, provides a
reasonable overall fit to the available data on L1544; the fit is
significantly worse for the non-magnetic model, in which the cloud
collapses promptly. The observed large peak column density for N$_2$H$^+$
and clear central depression for CCS favor the magnetically-retarded
collapse over the free-fall collapse. A relatively high abundance
of CCS is found in the magnetic model, resulting most likely from
an interplay of depletion and late-time hydrocarbon chemistry enhanced
by CO depletion. These initial results lend some support to the standard 
picture of dense core
formation in strongly magnetized clouds through ambipolar diffusion.
They are at variance with those of Aikawa et al. (2001) who considered
a set of models somewhat different from ours and preferred one in which
the cloud collapses more or less freely for L1544.
\end{abstract}


\clearpage

\section{Introduction}

Through a close interplay between theory and observation, a ``standard''
picture for the formation of isolated, low-mass stars has emerged
(Shu, Adams \& Lizano 1987). At the heart of this picture lie the
so-called ``dense cores'', studied extensively by Myers and coworkers
(e.g., Myers 1999). These dense cores are intimately associated with
star formation, with roughly half of them already harboring infrared
sources (Beichman et al. 1986). The other half are termed ``starless
cores'', and they are the focus of our investigation.

The starless cores are thought to be condensed out of strongly magnetized,
turbulent background clouds. Direct Zeeman measurements to date, as
compiled by Crutcher (1999), suggest that the field strength is close to
the critical value required for the molecular cloud support, after likely
geometric corrections (Shu et al. 1999). Core formation can be driven 
by either turbulence decay (Myers 1999) or ambipolar diffusion. We
shall focus on the latter, which has been studied quantitatively 
by many authors (e.g., Nakano 1979;
Lizano \& Shu 1989; Basu \& Mouschovias 1994; Ciolek \& Mouschovias 1994).
A general conclusion is 
that dense cores are formed on a time scale several times the free fall
time of the background.
This relatively long formation time should leave a strong imprint on
the chemistry of the cores.

The goal of our investigation is to predict the spatial distributions of
various molecular species at different stages of cloud evolution. For
this initial study, we will concentrate on CO, CS, CCS, NH$_3$,
N$_2$H$^+$ and HCO$^+$, which are often used to probe the physical conditions
of star-forming clouds. We take
into account the feedback of chemistry on the cloud dynamics, through
ion abundances, which regulate the ambipolar diffusion and thus
cloud contraction rate. Through a detailed comparison of the predicted
and observed chemical abundances of
starless cores, we seek to provide further support for the standard
picture of isolated star formation involving ambipolar diffusion.
A starless core well suited for such a purpose is the
dense core of L1544, for which an extensive data set is becoming
available (e.g., Caselli et al. 2001a,b).

The L1544 starless core is located in the eastern part of the Taurus
molecular cloud complex, at an estimated distance of 140 pc (Elias
1978). It is an elongated core that shows substantial infall motions
(up to $\sim 0.1$ km s$^{-1}$) on both large ($\sim 0.1$ pc)
and small ($\sim 0.01$ pc) scales, based on single-disk observations
of CO, $^{13}$CO, C$^{18}$O, CS, C$^{34}$S, HCO$^+$, and N$_2$H$^+$
(Tafalla et al. 1998) and interferometric observations of N$_2$H$^+$
(Williams et al. 1999). Interferometric observations of CCS by
Ohashi et al. (1999) reveal a ring-like structure, which shows
evidence for collapse as well as rotation. The low central emission
is most likely due to a depletion of CCS, judging from the fact that
dust continuum peaks inside the ring (Ward-Thompson, Motte \& Andre
1999). Indeed,
the column density distribution of the L1544 core is rather well
determined\footnote{See, however, Evans et al. (2001) who find that
a singular isothermal sphere cannot be ruled out from the
modeling of SCUBA data alone, when the lower temperature at the
core center relative to edge is taken into account.}, not only from the
dust continuum emission, but also from
absorption against a mid-infrared background (Bacmann et al. 2000).
The core has a small, high column (and volume) density central
plateau (of radius $\sim 2900$ AU), surrounded by an envelope in which
the column (and volume) density decreases rapidly outward, before
it joins rather abruptly with the background at a radius of order
$10^4$ AU (Bacmann et al. 2000). The
plateau-envelope structure has been interpreted in terms of the
ambipolar diffusion-driven dynamical evolution of a magnetized cloud 
(Li 1999; Ciolek \& Basu 2000), and the large column density contrast
between the central plateau and the background ($\sim 20$; Bacmann
et al. 2000) suggests that the L1544 core is rather evolved (Williams
et al. 1999), and may be on the verge of forming a star (or stellar
system). Further evidence for the core being in an advanced stage of
evolution comes from IRAM 30m observations of Caselli et al. (1999),
who found that CO is depleted by a factor of order $\sim 10$ at the dust
continuum peak. These authors also noted that the (likely) optically 
thin lines of D$^{13}$CO$^+$ and HC$^{18}$O$^+$ are double-peaked,
which may be another indication that these species are centrally
depleted and their emission comes mainly from an infalling {\it shell}.
The wealth of molecular line and dust continuum data, coupled
with the recent Zeeman measurement of magnetic field strength
(Crutcher \& Troland 2000) and the determination of magnetic
field direction from submillimeter polarization measurements
(Ward-Thompson et al. 2000), makes L1544 a Rosette Stone for
theories of dense core condensation leading to isolated low-mass star
formation.

Many of the dynamical and chemical features of L1544 are shared by
L1498, another quiescent, elongated starless core also located in 
the Taurus molecular complex. Lemme et al. (1995) studied the core in
detail, and found that the distributions of both C$^{18}$O and CS
are ring-like, indicating a central depletion of these two species.
The CO depletion factor is inferred to be of order $\sim 8$ or higher
(Willacy, Langer \& Velusamy 1998), comparable to that of
L1544, based on ISOPHOT 100 and 200 $\mu$m observations. Single-dish
and interferometric observations of CCS (Kuiper, Langer \& Velusamy
1996) show that its distribution is also ring-like. Moreover, the
NH$_3$ emission appears to be centrally
peaked, again similar to L1544. Perhaps most strikingly,
there is evidence for infall motions on a large ($\sim
0.1$ pc) scale in the double-peaked profiles of
both optically thick CS lines and the likely optically thin
C$^{34}$S(2-1) transition (Lemme et al. 1995); the latter could
come from an infalling shell, as in L1544 (Caselli et al. 1999).
The optically thin tracer N$_2$H$^+$ shows, however, a single
peaked spectrum (Lee, Myers \& Tafalla 2001), indicating that it is
centrally peaked, in contrast to C$^{34}$S. These similarities
to L1544 point to a relatively advanced stage of pre-protostellar
evolution for L1498, making it another valuable testing ground for
chemical and dynamical models of (low-mass) star formation.

Common to both sources is the ring (or shell)-like structure of CCS, 
a well-known ``early time'' species (Bergin
\& Langer 1997) tracing chemically ``young'' material of order
$10^5$ yrs or less. This prominent feature should provide a
strong constraint on models. Additional constraints should come
from the inferred large CO depletion near the core center as well as the
spatial extent and speed of the infall motions present in both L1544 and
L1498. The question we want to address is: can the standard picture
of dense core formation involving ambipolar diffusion satisfy these
chemical
{\em and} kinematic constraints simultaneously? A partial answer to
this question has been provided by Bergin \&
Langer (1997), who considered the chemical evolution of {\it a parcel} of
cloud material whose density increases with time in a prescribed way
(i.e., a standard time dependent but depth independent model).
They found that differential depletion of various molecular species onto
dust grains must be a key ingredient in understanding the differentiated
chemical structure observed in L1498.

Aikawa et al. (2001) went one step further, and followed the evolution
of molecular abundances of a dynamically collapsing starless core
{\it as a whole}, with an
emphasis on the spatial distribution. The study was based on the
self-similar solution of Larson (1969) and Penston (1969) and its
artificially delayed analogs. Among
the dynamical models adopted, they found that the undelayed Larson-Penston
model provides the best overall fit to the observational data on
CO, CCS and N$_2$H$^+$ for L1544. If this result is robust, its
implications would be far reaching.
It is well known that the Larson-Penston solution describes the collapse
of a cloud essentially on a free-fall time scale. The fact that it
matches observations better than its delayed analogs appears to
pose a serious challenge
to the standard picture of star formation involving strongly magnetized
cloud cores, which are formed through ambipolar diffusion on a time
scale much longer than the free-fall time scale. However, the large infall
velocity of the Larson-Penston solution, approaching 3.3 times the sound
speed, is clearly incompatible with those inferred for L1544 and other
starless cores, which are typically less than the sound speed
(Lee et al. 2001). Furthermore,
the best-fit model of Aikawa et al. under predicts the abundance of
N$_2$H$^+$
by a large factor of $\sim 20$. These discrepancies motivate us to
reexamine the
molecular evolution of starless cores using a more sophisticated dynamical
model, one that is coupled to the cloud chemistry.

The dynamical model we will adopt is that of Li (1999). It is a spherical 
model
of starless cores that takes into account the dynamic effects of magnetic
field
approximately. As envisioned in standard scenario of low-mass star
formation, the cores in the model evolve through ambipolar diffusion, and
their dynamics are coupled to chemistry through the abundances of
charged species. Previously, we (Shematovich, Shustov, \& Wiebe 1997; 1999)
have modeled the chemistry of dynamically evolving clouds in detail. The
chemical model will be combined with the dynamical model into a coupled
dynamical-chemical model of magnetized starless cores. An advantage
of the combined model is that it allows for a determination of not only
the time evolution of various
molecular species but also their {\em spatial} distributions {\em and}
kinematics. We describe the formulation of the combined model in
\S\ref{review}. In \S\ref{result}, we present representative model
results, focusing on the density distribution, velocity field and 
abundance distributions of
several commonly used molecular species. These results are compared with 
observations of L1544.
We conclude and discuss our main results and future refinements in
\S\ref{discuss}.

\section{Formulation of Coupled Dynamical-Chemical Model}
\label{review}

\subsection{Dynamical Model}

Ambipolar diffusion-driven evolution of strongly magnetized clouds
has been modeled by many authors, although a complete understanding
is still lacking. The numerical complexity of the problem
forced most workers to adopt either a quasi-static approximation
(e.g., Nakano 1979; Lizano \& Shu 1989) or a simplified disk-like
geometry (e.g., Basu \& Mouschovias 1994; Ciolek \& Mouschovias
1994). Some 2D axisymmetric dynamical models are available (Fiedler
\& Mouschovias 1993; Desch \& Mouschovias 2001), although
incorporation of a full-blown
chemical network into such models would be prohibitively expensive
computationally at present. A simpler approach
to the dynamical problem was pioneered by Safier, McKee \& Stahler (1997).
By retaining only the pressure component of the magnetic forces
they were able to retain a spherical geometry, which simplifies
the dynamics greatly. Indeed, they were able to obtain {\it analytic}
solutions to both core formation and collapse assuming further that
the thermal pressure is negligible and that the ions (and magnetic
field lines tied to them) move much more slowly than neutrals.
These last two assumptions have been relaxed in Li (1998) and the
resulting set of governing equations can be solved efficiently,
particularly in a Lagrangian form (Li 1999). The formulation
of the dynamical problem by Li (1999) is employed in our coupled
dynamical-chemical model.

For reference, we list below the equations governing the dynamic
evolution of a spherical, magnetized cloud (taken from Li 1999)
in dimensionless radius $\zeta$, mass $m$, density ${\hat \rho}$, 
velocity $u$, magnetic field strength $b$ and time $\tau$:
\begin{equation}
{\partial\zeta\over\partial m}={1\over\hat\rho\zeta^2}\,,
\label{eq09}
\end{equation}
\begin{equation}
{\partial u\over\partial\tau}=-{m\over\zeta^2}-\zeta^2
{\partial\over\partial m}\left({\hat\rho\over2\alpha_{\rm c}}
+{b^2\over2}\right),
\label{eq10}
\end{equation}
\begin{equation}
{\partial\over\partial\tau}\left({b\over\hat\rho\zeta}\right)=
{\partial\over\partial m}\left({1.4\over\nu_{\rm ff}}
{b^2\zeta^2\over\hat\rho^{1/2}}
{\partial b\over\partial m}
\right),
\label{eq_diff}
\end{equation}
\begin{equation}
u={\partial\zeta\over\partial\tau}.
\label{eq12}
\end{equation}
The density $\rho$ and field strength $B$ are scaled by their
initial values at the cloud center $\rho_{\rm c}$ and $B_{\rm c}$. The
time $t$ is measured in units of the initial free-fall time
\begin{equation}
t_{\rm ff,c}={1\over (4\pi G \rho_c)^{1/2} },
\label{t_ff}
\end{equation}
at the cloud center, and the velocity $V$ in units of the Alfv\'en speed
\begin{equation}
V_{\rm A,c}={B_{\rm c}\over (4\pi\rho_{\rm c})^{1/2}}.
\end{equation}
Scales for the radius $r$ and mass $M$ are then $r_{\rm c}=V_{\rm A,c}\,
t_{\rm ff,c}$
and $M_{\rm c}=4\pi\rho_{\rm c}r_{\rm c}^3$. There are two dimensionless
quantities in the governing equations. The first, $\alpha_{\rm c}$,
is the ratio of magnetic pressure to thermal pressure at the cloud center
at $t=0$. For simplicity, it will be taken to be unity, consistent
with the current
data on the magnetic fields in the interstellar clouds (Crutcher 1999).
The second is the magnetic coupling parameter $\nu_{\rm ff}$ defined as
the ratio of the local free-fall time (at a given density $\rho$) to the
magnetic field--neutral coupling time scale
\begin{equation}
\nu_{\rm ff}={\gamma \rho_{\rm i}\over
\sqrt{4\pi G\rho}},
\end{equation}
where $\rho_i$ is the total mass density of ions, and $\gamma$ is an
ion-neutral drag coefficient. An approximate value for the drag
coefficient of $\gamma=3.5\times 10^{13}$~cm$^3$~g$^{-1}$~s$^{-1}$
is given in Shu (1992), obtained assuming a mean molecular weight of
ions of $\sim 30$. We will adopt the above simple prescription for
the coupling parameter in this initial study, although a more
elaborate treatment taking into account of contributions from
charged dust grains is possible (e.g., Nishi, Nakano \& Umebayashi
1991; Li 1999) and will be implemented in the future.

\subsection{Chemical Model}

The chemical model used here has been described in detail by Shematovich
et al. (1997, 1999). Here we summarize the main features
briefly. We solve the equations of chemical kinetics describing
reactions in interstellar clouds of relatively low density
($\sim10^2-10^8$~cm$^{-3}$) and temperature ($T\sim10-100$~K).
Molecules are formed and destroyed both in gas-phase and on grain surfaces.
The dust grains are supposed to be well mixed with gas and contain
1\% of the cloud mass. We label different species, both neutral and
charged, by $i = 0,1,\ldots,M$, with $i = 0$ denoting electrons.
The evolving chemical composition
is described through the number densities $n_i^{\rm g}(r,t)$ in gas-phase
and $n_i^{\rm d}(r,t)$ on icy mantles. The exchange between gas and dust
is driven by accretion and desorption processes. In general, the
chemical abundances of all species are computed by integrating
the equations
\begin{equation}
{{\rm d}\over{\rm d}t}n_i^{\rm g}(r,t)=
\sum\limits_j\sum\limits_l K_{lj}^{\rm g}n_l^{\rm g}n_j^{\rm g}-
n_i^{\rm g}\sum\limits_jK_{ij}^{\rm g}n_j^{\rm g}-
k_i^{\rm acc}n_i^{\rm g}+k_i^{\rm des}n_i^{\rm d},
\label{chem1}
\end{equation}

\begin{equation}
{{\rm d}\over{\rm d}t}n_i^{\rm d}(r,t)=
\sum\limits_j\sum\limits_l K_{lj}^{\rm d}n_l^{\rm d}n_j^{\rm d}-
n_i^{\rm d}\sum\limits_jK_{ij}^{\rm d}n_j^{\rm d}+
k_i^{\rm acc}n_i^{\rm g}-k_i^{\rm des}n_i^{\rm d},
\label{chem2}
\end{equation}
where $K_{ij}^{\rm g}$ and $K_{ij}^{\rm d}$ are the gas-phase and
grain surface chemical reaction rates. The coefficients $k_i^{\rm
acc}$ and $k_i^{\rm des}$ define the accretion and desorption rates
for the $i$th component. These equations are integrated, along with
the dynamical equations, at each point of the flow. The computed
abundances are then used to determine the magnetic coupling parameter
$\nu_{\rm ff}$, and to obtain column densities for various commonly
used molecular tracers for comparison with observations.

\subsubsection{Gas-Phase Reactions}

We consider gas-phase reactions in the dense, dark regions of 
molecular clouds that are well shielded from the external 
UV radiation field. The reactions are initiated by galactic 
cosmic rays which ionize the most abundant species
with an adopted rate of $1.3\times 10^{-17}$ s$^{-1}$. The primary
ions that are formed react quickly with abundant hydrogen and CO
molecules, defining the main paths of chemical evolution.

We use the UMIST Rate95 file (Millar, Farquhar \& Willacy 1997).
The included elements are H, He, C, N, O, S, Si, Na, Mg, and Fe.
Our chemical reaction network is mainly limited to those species
containing no more than two atoms of trace elements (i.e. all
elements other than
H and He), with a few exceptions. One of these exceptions is 
C$_2$S, which has been mapped interferometrically in L1498
(Kuiper et al. 1996), L1544 (Ohashi et al. 1999),
and others. Its abundance and spatial distribution
provide a crucial constraint on models, as discussed by Aikawa
et al. (2001) and shown below. In
total, our gas-phase network contains about 2000 reactions between
59~neutral and 83~ion species.

\subsubsection{Grain Surface Reactions}

The formation of icy mantles due to accretion and desorption
has been taken into account. The accretion rate is given by
\begin{equation}
k_i^{\rm acc}=\pi a^2<v_i>S_in_{\rm gr}s^{-1},
\label{grain}
\end{equation}
where $a$ and $n_{\rm gr}$ are the grain radius (taken to be
10$^{-5}$ cm) and dust number density, $<v_i>$ is the
average thermal velocity of the species, and $S_i$
the sticking probability. We assume $S_i=0.3$ for all neutral
species except H$_2$ and He (Willacy, Rawlings \& Williams 1994). Two
desorption mechanisms are included. They are thermal evaporation and
cosmic ray induced impulsive thermal evaporation (Hasegawa \& Herbst 1993).
The dust temperature is
assumed to be 7~K everywhere, close to the value inferred by Evans et al.
(2001) and Zucconi, Galli \& Walmsley (2001) near the 
center of L1544. The values of the sticking probability and desorption
energies are uncertain. The sensitivity of our model to these
uncertainties will be explored elsewhere.

After being accreted onto the grain surface, the species are allowed 
to react with one another mainly through the hydrogen addition
reactions, with rates taken from Hasegawa, Herbst \& Leung (1992).
The surface products include saturated molecules (such as CH$_4$, 
H$_2$O and NH$_3$), homonuclear molecules (such as C$_2$, N$_2$, 
and O$_2$) and their hydrogenated forms, among
others. The accretion limit to the surface hydrogen addition chemistry
was taken into account through the first modification of rate equations
proposed by Caselli, Hasegawa, \& Herbst (1998). In addition to 
53 surface reactions (including molecular hydrogen formation),
we also consider the recombination of ions on grain surfaces. The
products of dissociative recombination are assumed to return into
the gas phase immediately. The effects of surface reactions 
on abundance distributions will be fully examined in a subsequent
paper. 

\subsection{Initial and Boundary Conditions}

We adopt as a starting point for our calculations an isolated
molecular clump in mechanical equilibrium, with a magnetic
pressure equal to the thermal pressure everywhere. The gas
temperature is assumed to be 10~K and is kept constant throughout 
the computations. The initial molecular hydrogen central density is set
to $n_{\rm H_2}=10^3$ cm$^{-3}$ for all models. Without ambipolar 
diffusion, the cloud would
remain in the equilibrium forever. At $t=0$ ambipolar diffusion is
switched on. The cloud evolution is then followed numerically
with reflection boundary conditions (i.e., zero spatial gradient)
at the cloud center and with a free pressure boundary at the cloud 
outer edge.

The initial chemical composition is assumed to be mostly atomic. Species
with non-zero initial abundances are listed in Table~\ref{iniab}. Their
abundances are given with respect to the number density of hydrogen
nuclei. All hydrogen is assumed to be in a molecular form initially.

\begin{table}
\caption{Species with Non-Zero Initial Abundances}
\begin{tabular}{ll}
\hline
\hline
Species&Abundance\\
\hline
H$_2$ & 0.5\\
He & 0.1\\
Na & 2.1($-$6)\\
Mg & 1.5($-$7)\\
Fe & 1.0($-$7)\\
C & 1.0($-$5)\\
N & 2.2($-$5)\\
O & 1.3($-$4)\\
CO & 4.0($-$5)\\
S & 8.0($-$8)\\
Si & 8.0($-$9)\\
\hline
\end{tabular}
\label{iniab}
\end{table}

\subsection{Numerical Method}

We divide the computational domain into spatial zones of equal mass. 
At each time step, we first advance the chemical model to evaluate
ion density $\rho_{\rm i}$ in each spatial zone. Equations
(\ref{eq09}), (\ref{eq10}), and (\ref{eq12}) are then solved to update
the velocity, radius of zone boundary and density. A new distribution
of magnetic field is obtained by solving the diffusion
equation (\ref{eq_diff}), with the magnetic coupling parameter
calculated from the ion densities at the beginning of the time
step.

The value of time step is computed from the usual magnetic
Courant-Friedrich-Lewy condition to ensure dynamic stability.
The chemical model is advanced over this time interval independently,
with its own algorithm and smaller time substeps that are
determined from the required abundance accuracy. We solve
the chemical kinetics equations with the standard LSODE package,
and terminate the computation when the central density reaches
$n_{\rm H_2}=10^6$~cm$^{-3}$. This density is close to the value
inferred at the center of L1544. At higher densities the
effects of charged dust grains on $\nu_{\rm ff}$ may become
dominant (Nishi et al. 1991), and such effects have not been
taken into account in the current version of the code.

\section{Representative Model Results}
\label{result}

The initial conditions for forming starless cores are not well constrained,
either observationally or theoretically. There is in principle a large
parameter space to be explored. In this initial study, we will
focus on two models that highlight the effects of magnetic field on
the dynamics and chemistry.
In the first model, we consider a standard, magnetic cloud with a mass
of 20 M$_\odot$ (StM+B). The second model is identical to the first except
that the magnetic field strength is set to zero (i.e., non-magnetic;
\hbox{StM--B}). For reference, we also consider the dynamics of
a model that is the same as the first
except that the magnetic coupling parameter is fixed at the canonical
value of $\nu_{\rm ff}=10$ instead of being computed self-consistently.
These models are summarized in Table~\ref{models}.

\begin{table}
\caption{Model Summary}
\begin{tabular}{lcl}
\hline
\hline
Model Name & $\nu_{\rm ff}$ & Magnetic Field \\
\hline
Reference & fixed at 10 & \hskip 1cm yes \\
StM+B & variable & \hskip 1cm yes \\
StM--B & not relevant & \hskip 1cm no \\
\hline
\end{tabular}
\label{models}
\end{table}

\subsection{Dynamics}

\begin{figure}
\epsscale{0.85}
\plotone{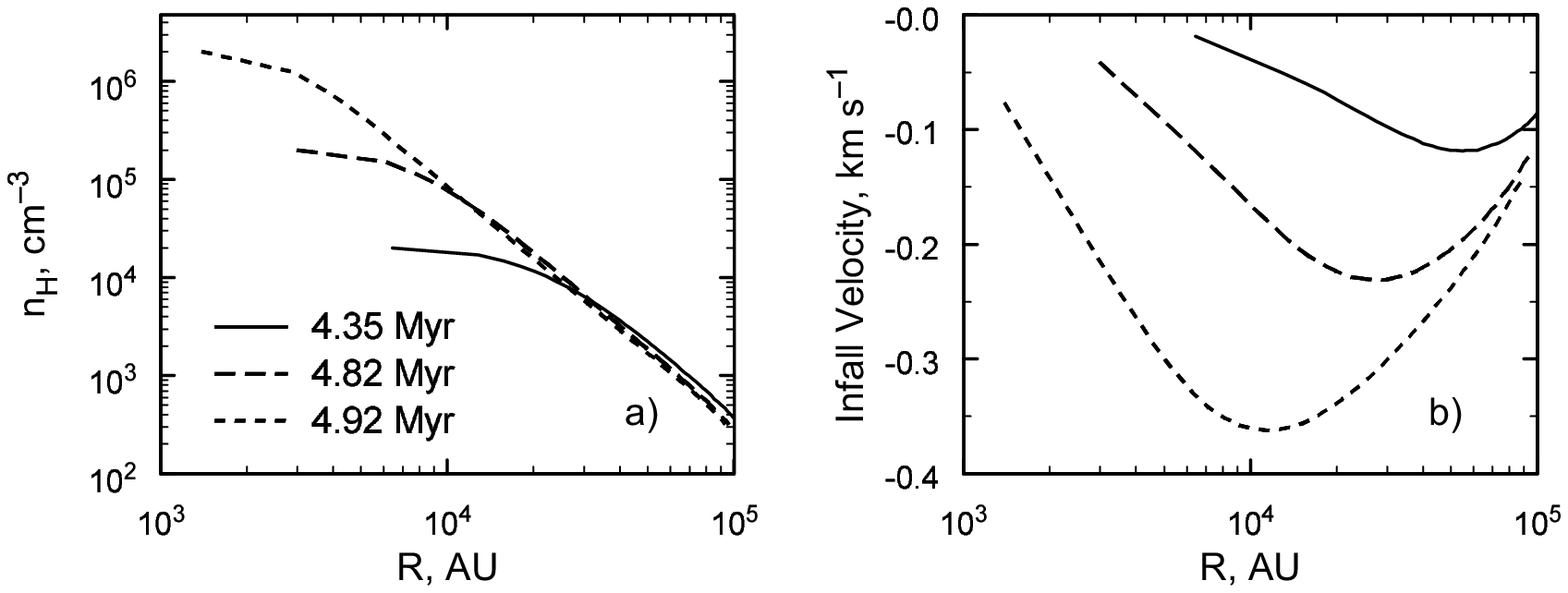}
\caption{Radial profiles of the number density of hydrogen nuclei
(a) and infall velocity (b) for the reference model at three times
when the central density reaches $10$, $10^2$ and $10^3$ times
its initial value respectively.\label{nomdyn}}
\end{figure}

\begin{figure}
\epsscale{0.85}
\plotone{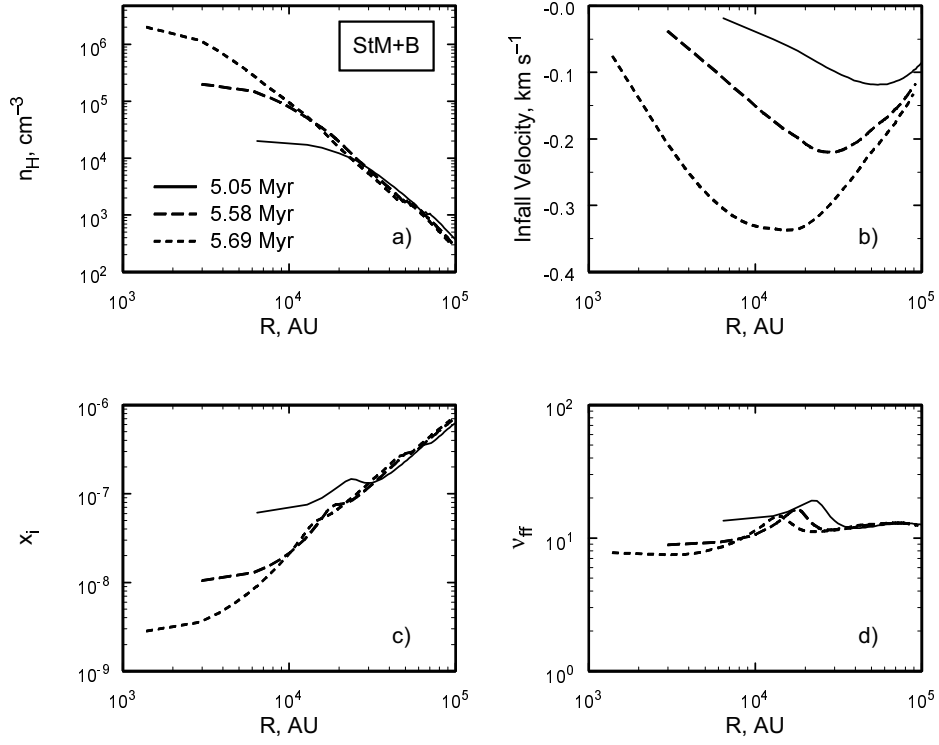}
\caption{Radial profiles of the number density of hydrogen nuclei (a),
infall velocity (b), ionization fraction (c), and magnetic coupling
parameter (d) for the standard, magnetic model.\label{stdyn}}
\end{figure}

\begin{figure}
\epsscale{0.85}
\plotone{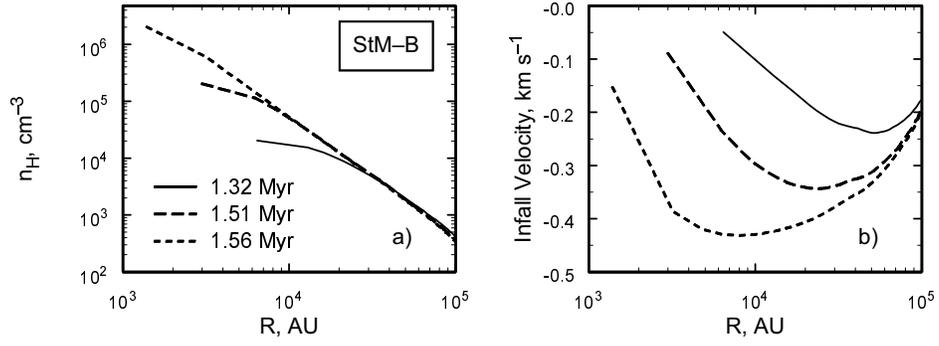}
\caption{Radial profiles of the number density of hydrogen nuclei (a)
and infall velocity (b) for the non-magnetic model.\label{stnobdyn}}
\end{figure}

The main dynamic quantities of the models (density distribution and
velocity field) are displayed in Figs.~\ref{nomdyn}--\ref{stnobdyn},
along with the ionization fraction and magnetic coupling parameter
for the StM+B model. As mentioned earlier, all models start
with a central density of $n_{\rm H_2}=10^3$~cm$^{-3}$, typical of
clouds in transition from being diffuse to dark. Note that when a
certain center density is reached, the difference in density profile
is modest, with that in the non-magnetic model somewhat peakier
than those in the magnetic models. The difference in the time it
takes to reach a given density is more pronounced. In
Table~\ref{tabtimes} we list the time intervals taken to
increase the central density by successive factors of 10. The
time for the first factor-of-10 increase, to a value of 10$^4$
cm$^{-3}$ characteristic of dense molecular cores, is $\sim 1$ Myr
for the non-magnetic model. It is substantially shorter than
those for the magnetic models ($\sim 4-5$ Myrs). The
times for subsequent 10-fold increases become shorter and shorter,
and the differences in time between the models also get smaller.
The time for the second (and third) factor-of-10
increase is longer for the magnetic models than for the non-magnetic
one by only a factor of $\sim 2.5$ (and $\sim 2$). In other words, after
formation a magnetic starless {\it dense core} typically lasts for
only a few free-fall times before stellar birth. 
{\it That the evolutionary
time scale becomes closer to that of a non-magnetic, free-falling
model at higher densities is an important feature of the
ambipolar diffusion-driven cloud evolution.
This characteristic cannot be captured by simply adding a constant
delay factor to a non-magnetic model}, as done for example in Aikawa
et al. (2001). The differences in time scale among the models will
have a decisive influence on the chemistry, with potentially observable
consequences.

Another potentially observable difference among the models is the
velocity field. From Figs.~\ref{nomdyn}--\ref{stnobdyn} we find
that the infall speed is typically a few tenths of 1~km s$^{-1}$,
comparable to the isothermal sound speed of the cloud ($a=0.188$~km
s$^{-1}$ at 10 K). As expected, the non-magnetic model collapses
right away, approaching a maximum speed $\sim 2.3$ times larger
than the sound speed $a$ by the time of a $10^3$-fold increase
in the central density. Even this maximum speed is substantially
lower than that in a Larson-Penston flow, 
which is 3.3 times the sound speed. The infall speed in
magnetized models increases more gradually, reaching a {\it maximum}
value of $\sim 0.35$ km s$^{-1}$ (or about twice the sound speed)
for the StM+B and reference model when the 
central density reaches $n_{\rm H_2}=10^6$ cm$^{-3}$, roughly the
density inferred for the central region of L1544 (Ward-Thompson et
al. 1999). Note that the density distribution flattens near the
center for all three models, more so in the magnetic models than
in the non-magnetic one. The radius of the central flat region
is of order $\sim 2-3\times 10^3$~AU for the StM+B model,
consistent with those deduced by Ward-Thompson et al. (1999)
and Bacmann et al. (2000). The
predicted peak H$_2$ column density of $\sim 10^{23}$~cm$^{-2}$
also agrees with the value inferred by Ward-Thompson et al.
(1999) within a radius of $\sim 900$~AU. However, the observationally 
inferred infall velocity for L1544
of order 0.1 km s$^{-1}$ (Tafalla et al. 1998; Williams
et al. 1999; Ohashi et al. 1999) is lower than predicted.
The discrepancy, although substantial, is less severe in our model
than in those of Aikawa et al. (2001). It could be further reduced
by taking into account the projection effects in a more realistic
flattened geometry and the effects of magnetic tension on cloud 
dynamics (Ciolek \& Basu 2000). 

\begin{table}
\caption{Time Intervals (in Myrs) for Central Density Increase}
\begin{tabular}{cccc}
\hline
\hline
H$_2$ density (cm$^{-3}$) & Reference Model& StM+B &StM--B\\
\hline
$10^3$--$10^4$ & 4.35 & 5.05 & 1.32 \\
$10^4$--$10^5$ & 0.47 & 0.53 & 0.19 \\
$10^5$--$10^6$ & 0.10 & 0.11 & 0.05 \\
\hline
\end{tabular}
\label{tabtimes}
\end{table}

\subsection{Ionization Fraction and Magnetic Coupling Parameter}

Before showing results on chemical abundances, we comment on the
ionization fraction $x_{\rm i}$ (defined as the ratio of the
number density of all ions to that of hydrogen nuclei) and
magnetic coupling parameter $\nu_{\rm ff}$, both of which are
shown in Fig.~\ref{stdyn} for the StM+B model. The fraction $x_{\rm
i}$ decreases with density, as expected. Its lowest value of $\sim
3\times 10^{-9}$ at the center of the cloud is somewhat higher
than that inferred for L1544 by Caselli et al. (2001b), although
the inference is model dependent. The ionization fraction yields
a magnetic coupling parameter $\nu_{\rm ff}$ within a factor of
two of the canonical value of 10.

\subsection{Chemical Abundances and Comparison with L1544}
\label{abund}

\subsubsection{Observationally Inferred Abundances in L1544}

We concentrate on the abundances of six species, including CO, CS,
CCS, HCO$^+$, NH$_3$ and N$_2$H$^+$. Other species will be considered
in a subsequent paper of the series. For CO, CCS, HCO$^+$, and
N$_2$H$^+$, the column density distributions have been inferred from
observations in L1544. Caselli et al. (2001b) found that CO is
heavily depleted near the center of L1544, by a factor up to 
$\sim 10$. Its column density distribution is flat or slightly
depressed near the center, with an average central value of
$\sim 1.5\times 10^{18}$~cm$^{-2}$, although large scatter exists
in the data (see
their Fig.~5). Both HCO$^+$ and N$_2$H$^+$ are centrally
peaked, with a peak column density of $\sim 10^{14}$ and $\sim 2
\times 10^{13}$~cm$^{-2}$ respectively. Ohashi et al. (1999) mapped
L1544 interferometrically in CCS. They found that CCS peaks in
a ring of radius $\sim 7,500$~AU. The peak CCS column density
is $\sim 4\times 10^{13}$~cm$^{-2}$, which is $\sim 1.4$ times
higher than the value at the center (Aikawa et al. 2001). Observational
data on NH$_3$ and CS appear to be less complete. Using the Nobeyama 45 m
telescope, Suzuki et al. (1992) observed L1544 in NH$_3$ and
derived an average column density of $2.7\times 10^{14}$cm$^{-2}$
within a region of radius $\sim 6,000$~AU, corresponding to a
beamwidth of $80^{\prime\prime}$. L1544 has been observed in CS
(Tafalla et al. 1998), although the peak value and spatial distribution
of its column density are not yet available.

\subsubsection{Standard Magnetic Model}

\begin{figure}
\epsscale{0.85}
\plotone{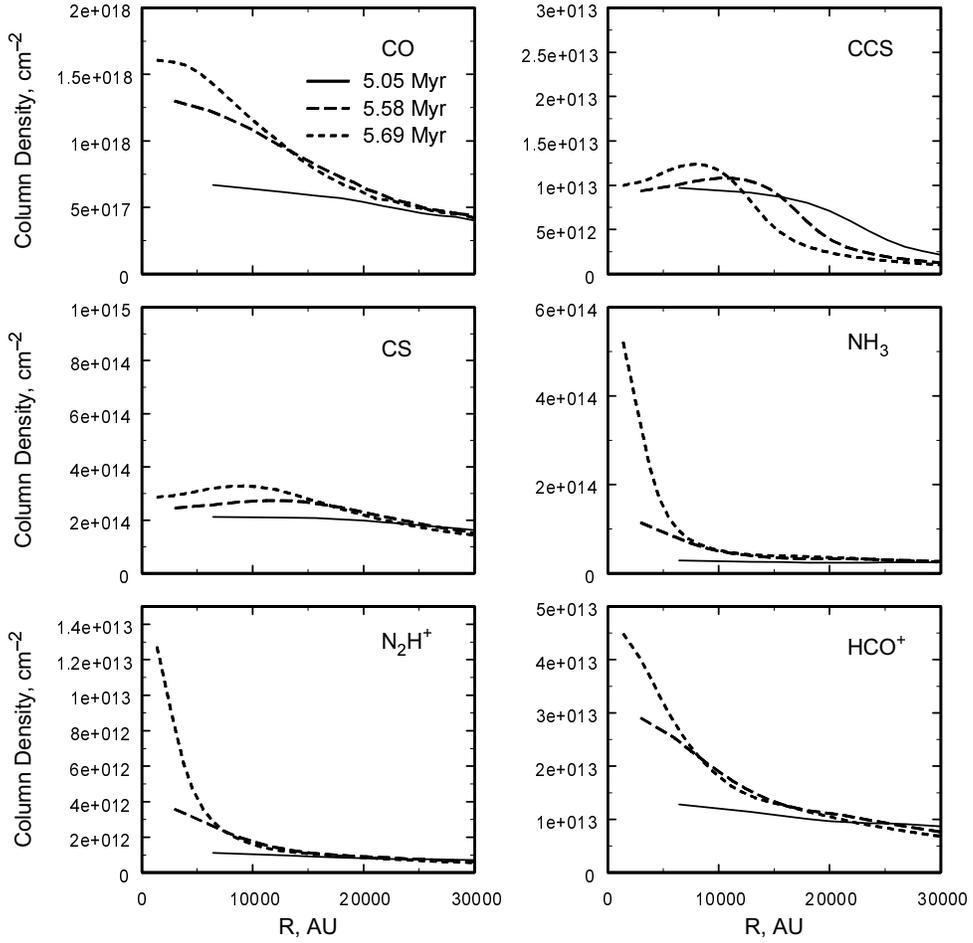}
\caption{Column density distributions of CO, CCS, CS, NH$_3$, N$_2$H$^+$,
and HCO$^+$ for the standard, magnetic model.\label{stb}}
\end{figure}

The column density distributions of the six species of interest for
the StM+B model, obtained by integrating the number densities along
the line of sight at various distances from line of sight through
the cloud center, are shown in
Fig.~\ref{stb}. We will concentrate on the last time (t=5.69 Myrs)
when the density distribution fits approximately that observed
(Ward-Thompson et al. 1999). It is immediately evident from the figure
that a central hole exists in
the distribution of CCS, with a radius ($\sim 7,000$~AU) and depth
($\sim 1.3$) that match almost exactly those observed (Ohashi et al.
1999). Note the differentiated chemical structure, with the centrally
peaked NH$_3$ and N$_2$H$^+$ surrounded by a CCS ring, which is
in turn embedded within a slightly broader CS ring\footnote{This is
not exactly what is observed in L1498, where CS appears to concentrate
near the inner edge of the CCS ring (Kuiper et al. 1996). The situation
with L1544 is unclear, although
there is some indication that a central hole exists in CS as well
(Tafalla 2001; priv. comm.).}. The predicted peak column density for
N$_2$H$^+$ of $1.3\times 10^{13}$~cm$^{-2}$ is within a factor of 2 of
the observed value ($\sim 2\times10^{13}$~cm$^{-2}$). The predicted
peak column density for NH$_3$ of $5.5\times 10^{14}$~cm$^{-2}$ is about
twice the value
$2.7\times10^{14}$~cm$^{-2}$ given by Suzuki et al. (1992). Averaging
the column density over a radius of 6,000 AU (the beamwidth) yields
a value $\sim 2\times 10^{14}$~cm$^{-2}$, which is in a closer
agreement with Suzuki et al.'s value. From Fig.~\ref{stb} we see that
HCO$^+$ is also centrally peaked, although less sharply than N$_2$H$^+$,
again in agreement with observation (see Fig.~5 of Caselli et al. 2001b). 
Its predicted peak column density
of $4.5\times 10^{13}$~cm$^{-2}$ is about a factor of two lower than
observed ($\sim 10^{14}$~cm$^{-2}$). The CO column density in this model
shows a relatively flat distribution near the center, with a maximum
value of $1.6\times 10^{18}$~cm$^{-2}$, almost identical to the
observed value. Its spatial distribution appears to be consistent with
that inferred by Caselli et al. (2001b) within errors. The model
predicts that the column density distribution of CS should have a
shallow depression near the center. Data for testing this
prediction are not yet available. We conclude that, within a factor of
two or so, the predictions of the StM+B model on the abundances of CO,
N$_2$H$^+$, HCO$^+$ and NH$_3$ are consistent with the data available
on L1544.

The abundance and spatial distribution of CCS deserve special attention.
We have mentioned the close fit to the observed spatial distribution
and will now concentrate on abundance. The CCS molecule is a well-known
``early time'' species. Its abundance
should decrease, according to Bergin \& Langer (1997), rapidly after
a few times $10^5$ years, as a result of neutral atomic carbon being
converted into CO (see also Suzuki et al. 1992 and Aikawa et al. 2001).
One might expect to find little CCS in the magnetic cloud of the StM+B
model, which reaches the observed state after nearly 6 million years
of evolution. However, this turns out not to be the case, as shown in
Fig.~\ref{stb}. The predicted
peak CCS column density is $1.3\times 10^{13}$~cm$^{-2}$, about a
third of the observed value ($\sim 4\times 10^{13}$~cm$^{-2}$). The
agreement would be closer, if one takes into account of likely
projection effects. The CCS ring is seen almost edge-on, judging from
its large aspect ratio ($\sim 3$; Ohashi et al. 1999). The
nearly edge-on geometry implies a smaller peak column density {\it
perpendicular to the ring}, by a factor of $\sim 3$. It turns
out that CCS plays a crucial role in discriminating models. We will
postpone a more detailed discussion of this molecule to
\S~\ref{ccs}. Here, we simply note that the StM+B model appears to
reproduce all abundance data on L1544 reasonably well. It is the 
``standard'' model against which other models will be compared. 

\subsubsection{Free-Falling Non-Magnetic Model}

\begin{figure}
\epsscale{0.85}
\plotone{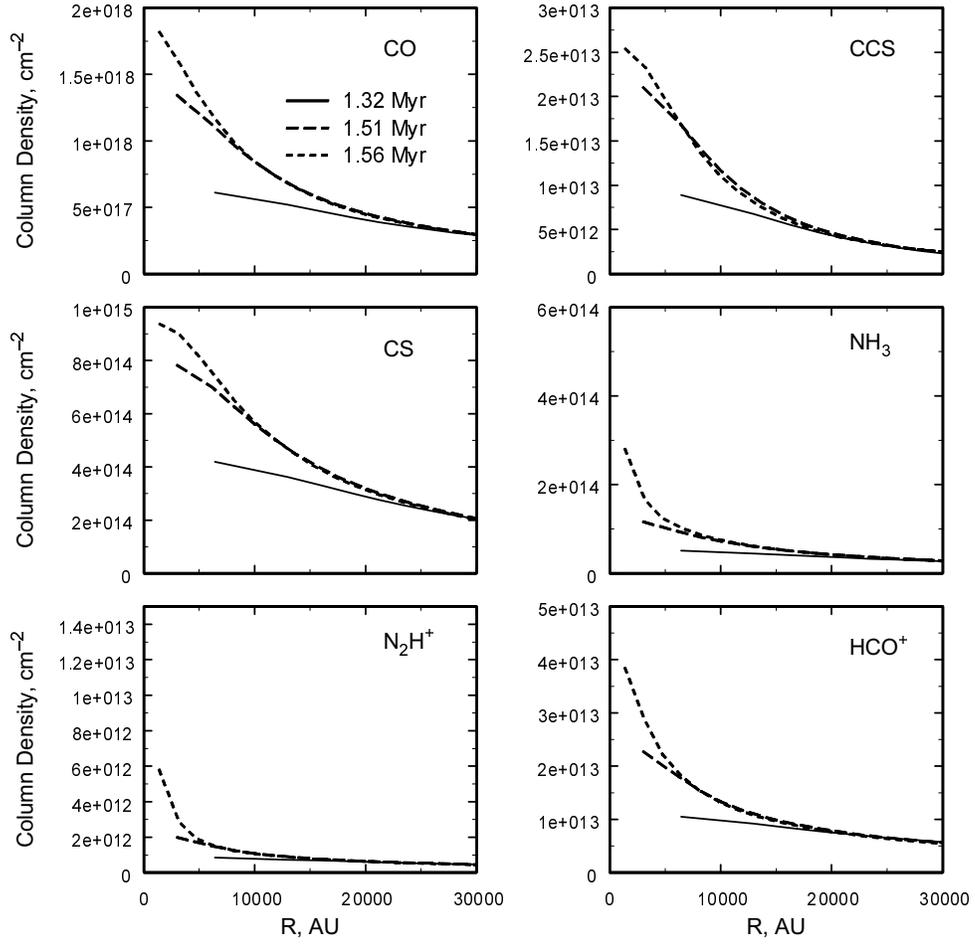}
\caption{Column density distributions of CO, CCS, CS, NH$_3$, N$_2$H$^+$,
and HCO$^+$ for the non-magnetic model.\label{stnob}}
\end{figure}

Without magnetic support, the cloud considered in the previous subsection
collapses promptly, reaching the observed state in 1.56 million years.
This faster dynamic evolution affects the chemical abundances, as shown
in Fig.~\ref{stnob}. The effects on CO, NH$_3$, N$_2$H$^+$ and HCO$^+$
are relatively modest. The peak column density of CO differs little from
that of the standard model. Its spatial distribution appears more
centrally peaked, however, which is more difficult to reconcile with
observation.
The distributions of N$_2$H$^+$, NH$_3$ and HCO$^+$ are centrally peaked,
as in the standard model. Compared with the standard model,
the peak value for N$_2$H$^+$ is down by a factor of $\sim 2$, making
it significantly below the observed value (by a factor of $\sim 4$). The
column density of NH$_3$ is also down by a similar factor of $\sim 2$, 
making the fit to the value inferred by Suzuki et al. (1992) worse. The 
column density of HCO$^+$ is lower by a modest factor of $\sim 20\%$.

The difference in CS and CCS is more pronounced. Unlike in the standard model,
the column density distributions of CS and CCS in the non-magnetic model
are centrally peaked instead of showing a central depression. While the
observational situation with CS is still unclear at present, a central peak
in CCS is definitively ruled out. One potentially desirable feature of the
non-magnetic model is that the peak column density of CCS is about twice
the value in the standard model, making it closer to the observationally
inferred value without projection corrections. However, its overall fit to
all available
data is significantly worse than that of the standard model.
This result lends some support the standard picture
of low-mass star formation involving core formation through ambipolar
diffusion over a time scale substantially longer than free fall. The peak
value and spatial distribution of the column density of CS would further
constrain the models but are not yet available. They are urgently
needed.

We note that our non-magnetic model is similar to the undelayed model 
of Aikawa et al. (2001), except that we have included some surface 
reactions and adopted a smaller sticking probability. 

\section{Discussion and Conclusions}
\label{discuss}

\subsection{Magnetic Cloud Support and CCS Abundance}
\label{ccs}

The success of the standard, magnetic model in fitting the data on L1544
hinges to a large extent on the relatively high abundance predicted for
CCS. This result is somewhat surprising in view of the fact that CCS
is a well-known ``early time'' species, whose abundance should decline rapidly
after a few times $10^5$ years according to Suzuki et al. (1992), Bergin
\& Langer (1997), and Aikawa et al. 2001). Part of the reason for the
persistently
high CCS abundance in the standard model
even after some 5.69~Myrs of evolution (the time it
takes the cloud to reach the
observed state) is that we adopted an initial cloud density of $n_{\rm
H_2}=10^3$~cm$^{-3}$, which is lower than those of Bergin \& Langer
($10^{3.5}$~cm$^{-3}$) or Suzuki et al. and Aikawa et al. ($10^4$~cm$^{-3}$).
As a result, the bulk of the $5.69$~Myrs is spent in the ``pre-dense
core'' phase when the central density remains below $10^4$~cm$^{-3}$ (see
Table~3). Only a small fraction ($\sim 10\%$) of that time is spent at
densities above $10^4$~cm$^{-3}$. That a cloud spends a relatively short
time (a few times the free-fall time) at high densities after forming a
magnetically supercritical core is a general feature of ambipolar
diffusion-driven evolution. It helps to keep the CCS abundance higher
than one would expect based on the total time of cloud evolution.

The fractional abundances of CCS are shown in Fig.~6 as a function of
radius for both the magnetic and non-magnetic models. Note the ``humps''
on the distributions of CCS abundance in the magnetic model, which
are not apparent in the non-magnetic model. The exact origin of the humps
is unclear. We suspect that they are related to the depletion of CO
in the central high density region, similar
to the ``depletion'' peak proposed by Ruffle et al. (1997) for
HC$_3$N. Ruffle et al. (1997, 1999) showed that
C$_2$H and HC$_3$N in dense, cold cores are characterized
by a secondary ``late-time'' maximum in their fractional abundances.
The second peak is thought to be mainly caused by the increased depletion
of CO, which allows C$^+$ ion to react more readily with
H$_2$ than with oxygen bearing species. The net result is
an increase in the rates of production of CH and
other carbon-bearing molecules (without oxygen). It was found that
high late-time C$_2$H and HC$_3$N maxima are achieved only when
the freeze-out time scale is long compared to the chemical time scale.
Because C$_2$H and other (neutral and ionized) late-time hydrocarbons
are the precursors of CCS formation in the reactions with neutral and
ionized sulfur through
\begin{equation}
{\rm {C_2H,C_2H_2,C_2H_3, ...} + S^{+} \rightarrow HCCS^{+} + ...., }
\end{equation}
\begin{equation}
{\rm {C_2H_2^{+},C_2H_3^{+},C_2H_4^{+}, ...} + S \rightarrow HCCS^{+} + ....,}
\end{equation}
\begin{equation}
{\rm HCCS^{+} + e \rightarrow CCS + H, }
\end{equation}
CCS should behave in a way similar to the C$_2$H and HC$_3$N. The
competition between depletion and late-time hydrocarbon chemistry
enhanced by CO-depletion appears to be the most likely cause of the
humps on the radial profiles of CCS abundance.

In any case, the humps have apparently kept the CCS abundance of the
magnetic model close to, and in some regions exceeding, that of the
non-magnetic model. They are largely responsible for the
higher-than-expected value of the CCS column density in the magnetic
model. In addition, the decline of CCS abundance outside the hump,
coupled with a steep decrease in the hydrogen number density with
radius, may explain the observed rapid decline of the CCS column
density outside the ring (Ohashi 2001; priv. comm.).

\subsection{Depletion of Molecules onto Dust Grains}

\subsubsection{Depletion of CO and Other Species}

In the standard model, CO is heavily depleted in the central high
density region.
The depletion is most clearly seen in the first panel of Fig.~7,
where the fractional abundance is plotted as a function of radius.
The abundance is significantly below the canonical value of $4.7\times
10^{-5}$ inside a radius of $\sim 10^4$~AU, by a factor up to $\sim 30$.
The size of the depletion region is roughly consistent with that
deduced by Caselli et al. (2001b; their Fig.~2). The depletion is less
evident in the CO column density distribution (see Fig.~4), which is
flat near the center. The ratio of CO and hydrogen in peak column
density is a factor of $\sim 5$ below the canonical value. This
depletion factor is about half of the value ($\sim 9$) inferred by
Caselli et al. (2001b). Their inference is based on the observations
of C$^{17}$O, C$^{18}$O, and dust emission (Ward-Thompson et al.
1999). In particular, an dust opacity of 0.005~cm$^2$g$^{-1}$ was
adopted to convert the dust continuum flux into a hydrogen column
density. At densities as high as $10^6$~cm$^{-3}$, an opacity of
$\sim 0.01$~cm$^2$g$^{-1}$ may be more appropriate (Ossenkopf \&
Henning 1994), which would lower the estimate of the hydrogen
column density by half and bring a closer agreement between the model
prediction
and observation. Alternatively, a flattened geometry, as observed for
L1544, and/or a somewhat higher CO adsorption energy (see Aikawa et al.
2001) may enhance the CO depletion in the column density distribution.

Besides CO, there are other species that are strongly depleted in
the high density central region. These include CS, CCS and to a
lesser extent HCO$^+$, as can be seen from Fig.~7. The nitrogen
bearing species, N$_2$H$^+$ and NH$_3$, are on the other hand hardly
depleted; if anything, their abundances are slightly enhanced near
the center. This differential depletion pattern is in agreement
with previous findings (e.g., Bergin \& Langer 1997; Aikawa et al.
2001). It may have profound effects on the molecular line profiles
used to probe cloud kinematics.

\subsubsection{Effects of Depletion on Line Profiles}

An interesting observational fact about L1544 is that some optically
thin lines in L1544, such as CCS (Ohashi et al. 1999) and HC$^{18}$O$^+$
(Caselli et al. 2001b), are double-peaked. As noted by Caselli et
al. (1999, 2001b), the double-peaked optically thin lines can be
explained in a collapsing cloud {\it provided} that the line-emitting
molecules are strongly depleted near the center. We have seen from
Fig.~7 that CCS and, to a lesser extent, HCO$^+$ (and thus HC$^{18}$O$^+$)
are indeed centrally depleted in our standard model. A unique
strength of the coupled dynamical-chemical model is that it allows for a
simultaneous determination of the spatial distribution of a given
species and the velocity field, the two ingredients for line profile
modeling. In a subsequent paper, we will produce synthetic line profiles
for both optically thin and thick lines (such as CS and HCO$^+$; Tafalla
et al. 1998), which should enable us to constrain the models of core
formation leading to star formation using both the spatial {\it and} 
kinematic information of molecular species.

\subsection{Conclusions}

We have developed a coupled dynamical and chemical model for the starless
cores of strongly magnetized molecular clouds. The coupling is
achieved through ions, the abundances of which are determined
self-consistently from a chemical network. The ionic abundances control
the time scale of cloud dynamic evolution via ambipolar diffusion.
We have concentrated on a representative model in which an isolated magnetic
cloud increases its central number density by a factor of $10^3$ to
$n_{\rm H_2}=10^6$~cm$^{-3}$ in 5.69 million years, with an eye on
explaining the observational data on the chemical abundances of the
well-studied starless core L1544.
We find that the predicted peak values of CO, CCS, N$_2$H$^+$ and
HCO$^+$ column densities are within a factor of two or so of those
inferred for L1544 from observations. The spatial distributions
of these species are also consistent with those observed. An
alternative, non-magnetic model was also considered for comparison.
With the same initial conditions as in the magnetic model, the 
non-magnetic cloud collapses promptly, reaching the observed state 
in merely 1.56 million years. It produces a worse overall fit to 
the available data on L1544. In particular, the column density 
distribution of CCS predicted by the model is centrally peaked, 
which is not observed.

We conclude that our initial results of modeling lend some support 
to the standard picture of dense core formation out of strongly 
magnetized molecular clouds involving ambipolar diffusion over 
several dynamic times. There are, however, a number of uncertainties
to which the results may be sensitive to, including the sticking
probability and adsorption energies for various species. A parameter 
survey is needed to firm up the conclusion.

\subsection{Future Refinements}

There are several aspects of the coupled model that we wish to improve
upon in the future. These include both dynamics and chemistry. Even
though our spherical model captures the essence of the ambipolar
diffusion-driven cloud evolution, the effects of magnetic tension
(which tend to flatten a cloud) cannot be treated. Indeed, the opposite
extreme, a disk-like geometry, may describe the observed elongated mass
distribution in L1544 better. It should be straightforward to extend
the model to this geometry, which would allow for the inclusion of
rotation whose presence has been inferred for L1544 by Ohashi et al.
(1999) based on
CCS data. On the chemistry side, a major uncertainty is the adsorption
energies for CO and other species. Different sets of adsorption
energies have been considered by Aikawa et al. (2001), and a similar
parameter study is needed for our model. Also uncertain are the
probability of molecules sticking onto dust grains and grain surface
processes, and an exploration of different cases is desirable.
Another area of improvement would be the treatment of
magnetic coupling coefficient, by including the effects of charged
dust grains (especially small ones) and external UV radiation field.

\acknowledgments

The authors are grateful to P. Murphy and B. Turner for access to
NRAO computing facilities, and to P. Caselli and the referee
for helpful comments. This research is supported in part by
RFBR Grant 01-02-16206.

\clearpage

\clearpage

\clearpage

\clearpage

\clearpage

\clearpage

\begin{figure}
\epsscale{0.85}
\plotone{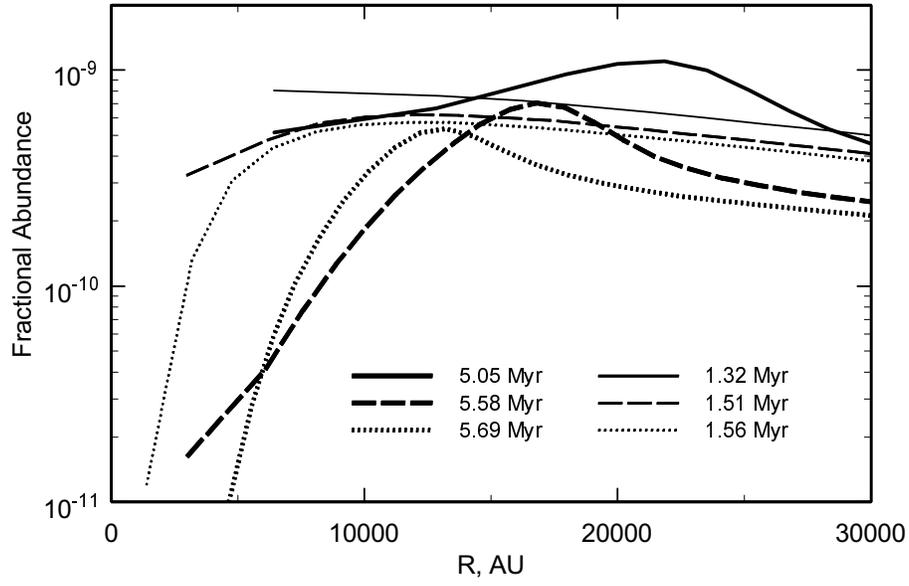}
\caption{Radial profiles of the fractional abundances of CCS for both
the magnetic (heavy lines) and non-magnetic (light lines) models.
\label{ccsabund}}
\end{figure}

\clearpage

\begin{figure}
\epsscale{0.85}
\plotone{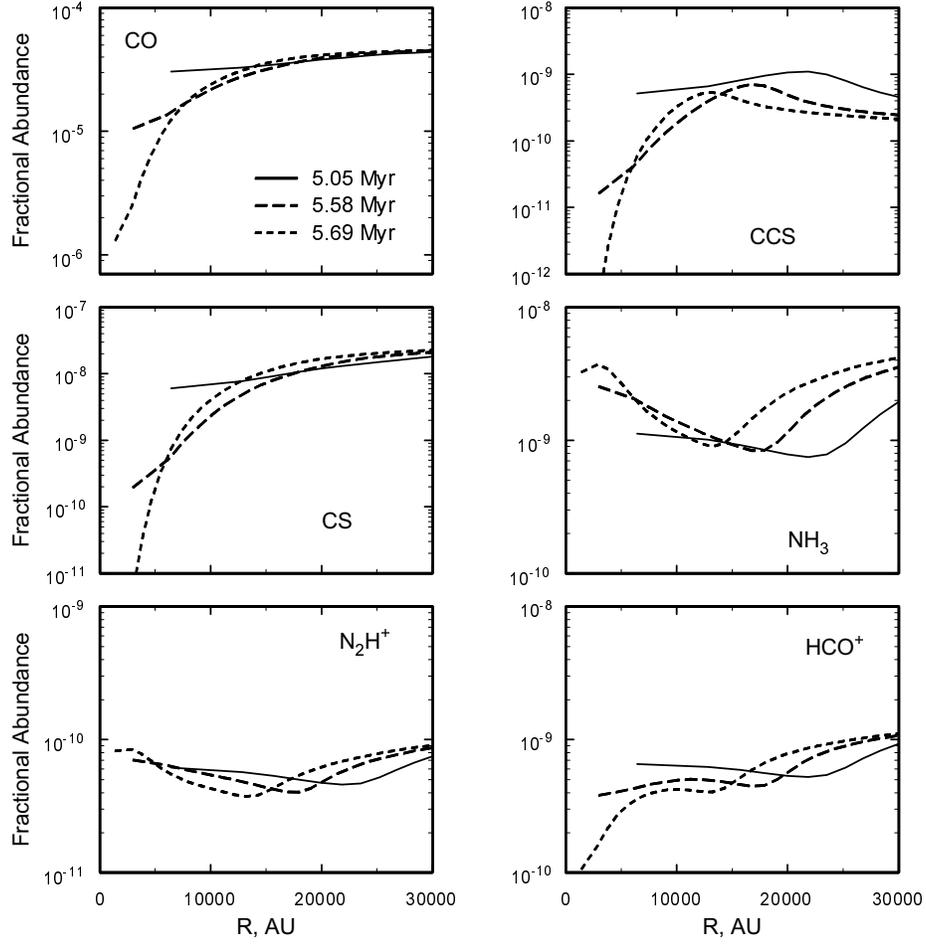}
\caption{Radial profiles of the fractional abundances of CO, CCS, CS,
NH$_3$, N$_2$H$^+$, and HCO$^+$ for the standard, magnetic model.
\label{lob}}
\end{figure}

\end{document}